\documentclass[dvips]{acta}
\usepackage{supertabular,lscape,epsfig}
\usepackage{amssymb}
\usepackage{amsmath}

\SetPages{0}{0}

\SetVol{58}{2008}

\begin{document}

\begin{Titlepage}
\Title{The Qatar Exoplanet Survey}
\Author{K.A.~Alsubai$^1$,
        N.R.~Parley$^2$,
        D.M.~Bramich$^1$,
        K.~Horne$^2$, 
        A.~Collier~Cameron$^2$,
        R.G.~West$^3$,
        P.M.~Sorensen$^4$,
        D.~Pollacco$^{5,6}$,
        J.C.~Smith$^7$
        and O.~Fors$^1$}
       {$^1$Qatar Environment and Energy Research Institute, Qatar Foundation,
                       Tornado Tower, Floor 19, P.O. Box 5825, Doha, Qatar \\
        e-mail: kalsubai@qf.org.qa \\
        $^2$SUPA, School of Physics and Astronomy, University of St Andrews, North Haugh, St Andrews, Fife, KY16 9SS, UK \\
        $^3$Department of Physics and Astronomy, University of Leicester, Leicester, LE1 7RH, UK \\
        $^4$Nordic Optical Telescope, Apartado 474, E-38700 Santa Cruz de la Palma, Santa Cruz de Tenerife, Spain \\
        $^5$Department of Physics, University of Warwick, Coventry, CV4 7AL, UK \\
        $^6$Astrophysics Research Centre, School of Mathematics \& Physics, Queens University, University Road, Belfast,
            BT7 1NN, UK \\
        $^7$Hidden Loft Observatory, Tucson, AZ 85755, USA}

\Received{Month Day, Year}
\end{Titlepage}

\Abstract{
The Qatar Exoplanet Survey (QES) is discovering hot Jupiters and aims
to discover hot Saturns and hot Neptunes that transit in front of
relatively bright host stars. QES currently operates a robotic
wide-angle camera system to identify promising transiting exoplanet
candidates among which are the confirmed exoplanets Qatar 1b and 2b.
This paper describes the first generation QES instrument, observing
strategy, data reduction techniques, and follow-up procedures. The QES
cameras in New Mexico complement the SuperWASP cameras in the Canary
Islands and South Africa, and we have developed tools to enable the QES
images and light curves to be archived and analysed using the same
methods developed for the SuperWASP datasets. With its larger aperture,
finer pixel scale, and comparable field of view, and with plans to
deploy similar systems at two further sites, the QES, in collaboration
with SuperWASP, should help to speed the discovery of smaller radius
planets transiting bright stars in northern skies.
}
{instrumentation - surveys - planetary systems}

\section{Introduction}

The surprising existence of short-period ($\sim$4~day) Jupiter-mass
extra-solar planets (termed ``hot Jupiters''), confirmed by radial
velocity measurements in the last decade, has shown us that planetary
systems exist in patterns unlike that of our own Solar System. The class
of hot Jupiter planets ($P<10$~d and $M\sin{i}<10\,M_{J}$) makes up
$\sim$35\% of the planets discovered to date, and $\sim$0.7\% of 
the transiting planets from the {\it Kepler} space mission host such
a companion (Dong \& Zhu 2013).
Given a fortuitous geometric alignment, an extra-solar planet may be
observed to transit the host star as viewed from the Earth. Such a
planetary transit is characterized by a small
decrease in the observed brightness of the host star that repeats at the
orbital period of the extra-solar planet. The probability that a typical
hot Jupiter transits its host star is $\sim$10\%
(Horne 2003), and hence, being conservative, $\sim$1 in 1400 stars
will host a transiting extra-solar hot Jupiter.

Five years after the discovery of the first extra-solar planet around
a sun-like star (Mayor \& Queloz 1995), the extra-solar planet in
orbit around HD~209458 was found to transit the stellar disk
(Charbonneau et al. 2000, Henry et al. 2000).
This hot Jupiter was already known to have
$M_{p}\sin{i}=0.69\pm0.05\,M_{J}$\footnote{$M_{J}$ is Jupiter's mass,
$M_{p}$ is the planet's mass and $i$ is the orbital inclination.}
from the radial velocity measurements (Mazeh et al. 2000). Also,
the spectral type, and hence the mass and radius, of the host star were
already known. Consequently, the modelling of the two observed transit
events allowed the measurement of the orbital inclination, which in turn
allowed the true mass of HD 209458b to be calculated.
Charbonneau et al. (2000) measured $i=87.1\pm0.2$ degrees,
implying that $M_{p}=0.69\pm0.05M_{J}$. They also measured
$R_{p}=1.27\pm0.03R_{J}$ from the transit fit (where $R_{p}$ is the
planet radius).

The importance of this result lies in the fact that for the first time
the mass and radius of an extra-solar planet had been measured, not just
a lower limit on the mass. Before this discovery, the radii of the
extra-solar planets were unknown and hence their average densities were
unknown. The average density derived for HD~209458b was $\sim0.38$
g/cm$^{3}$, significantly less than the average density of Saturn ($0.7$
g/cm$^{3}$), the least dense of the Solar System gas giants. This was
proof that HD~209458b must be a gas giant rather than a rocky
(terrestrial) planet, lending weight to the term hot Jupiter. Average
density was not the only important quantity that could be calculated for
an extra-solar planet for the first time. Other such quantities included
surface gravity and effective temperature.

\begin{figure}[htb]
\includegraphics[width=\textwidth,height=\textheight,keepaspectratio]{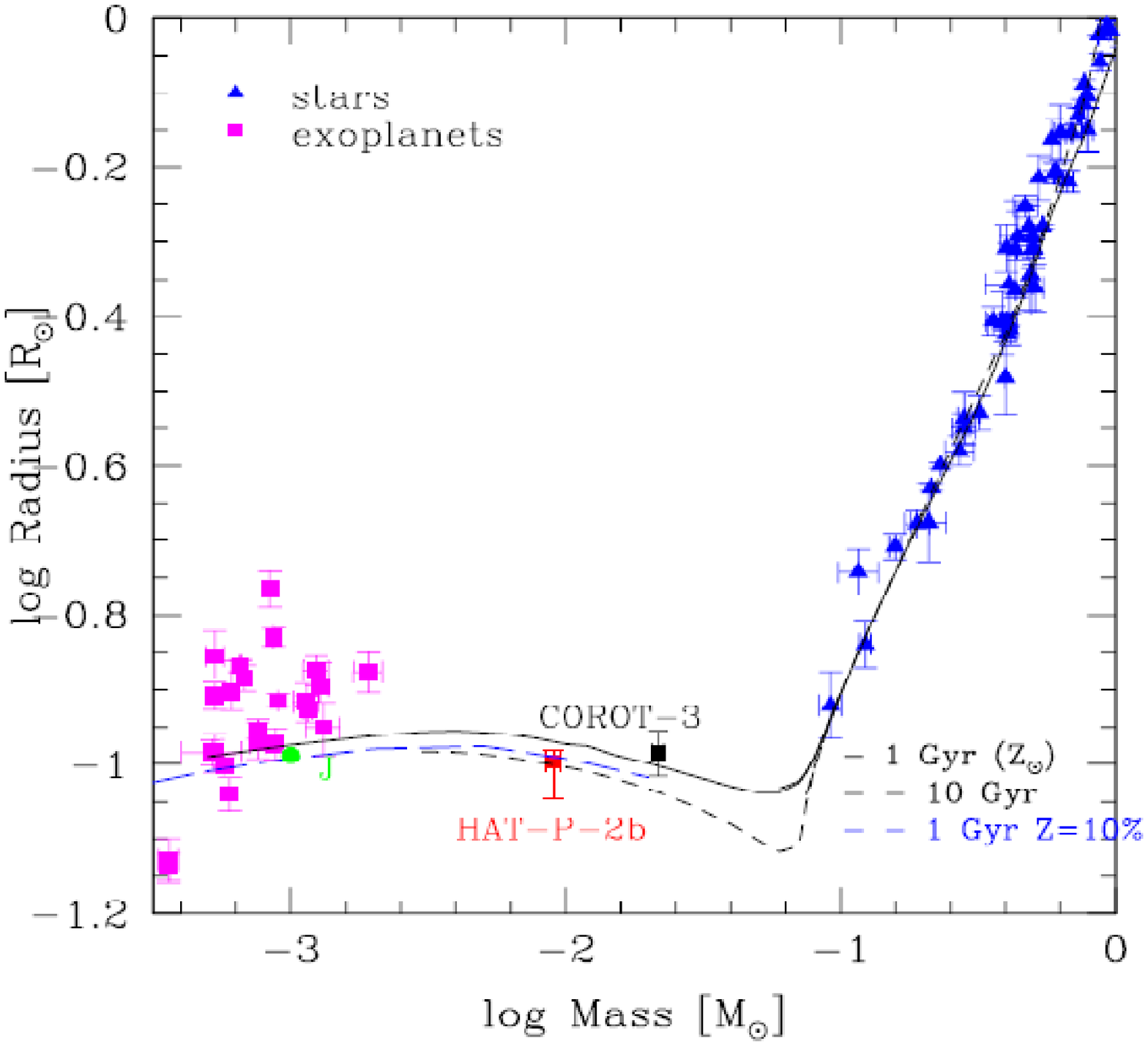}
\FigCap{Mass-radius relationship from the stellar to the planetary regime. The
        (black) solid and short-dashed lines correspond to models with solar
        composition, for two isochrones. The long-dashed line corresponds to an
        object with a $Z = 10$\% mass fraction of heavy elements. The
        observationally-determined values of HAT-P-2b and Corot-3b are indicated.
        Reproduced from Chabrier et al. (2009).}
\label{fig1}
\end{figure}

Since the discovery of the transiting nature of HD~209458b, some 424
transiting planets have been confirmed with periods from 0.18 to 904
days\footnote{See exoplanet.eu}. This class of extra-solar planet now
makes up a significant fraction of the 1047 known extra-solar planets to
date. The statistics provided from these systems are helping to pin down
the mass-radius relationship for these planets (Figure~\ref{fig1} reproduced
from Chabrier et al. 2009). They are also providing a challenge to
theories of planetary structure and evolution in order to explain the
observed radii (Liu et al. 2008), which depend on stellar
type, orbital distance, planet mass and age. The results from the
modelling of planetary structure have implications for the planetary
formation theories, especially in discriminating between the
core-accretion model (Ida \& Lin 2004) and the gravitational
instability scenario (Perryman 2000). The importance of
improving extra-solar planet statistics is paramount for advancing these
theories, and for helping to determine the exact definition of what
constitutes a planet in the transition regime between brown dwarfs and
planets.

The first major jump in the discoveries of transiting extra-solar
planets arrived with the advent of ground-based wide-field surveys of
bright stars. Many projects searching for transiting exoplanets have
preceded QES. Among those that have met with success are SuperWASP, HAT,
XO, TrES and KELT. These pioneering surveys pave the way for new
projects such as QES, since they have tackled and solved many hardware
and software issues.

A prototypical ground-based exoplanet search project
is the Wide-Angle Search for Planets (SuperWASP) survey
(Pollacco et al. 2006). The SuperWASP survey employs eight camera units
attached to a single robotic mount, where each camera uses an E2V
$2048\times2048$ pixel professional CCD backing a 200mm f/1.8 Canon
lens. Each unit has a field of view (FoV) of 60 square degrees at a
scale of 14.2 arcsec per pixel, giving a field coverage of $\sim$500
square degrees, which is imaged once per minute. SuperWASP targets stars
in the brightness regime from 8-13$^{th}$ magnitude,
leading to the discovery of some 80 transiting exoplanets to date.

Currently there is a gap in the magnitude range of stars being surveyed
for transiting planets, and this is the range from 12-15$^{th}$~mag
where SuperWASP targets become too faint for their instrumentation, and
stars are too bright for the deeper surveys, like OGLE
(Udalski et al. 2002), which start at 15$^{th}$ mag. Exploration
of this range is important, because it increases the sampling volume for
intrinsically faint K and M-dwarf stars, whose smaller radii facilitate
the detection of transits by small planets. QES is designed to
fill this gap and detect transiting planets in the range 10-14$^{th}$
magnitude by constructing survey equipment targeted at this magnitude
range.

We are planning to deploy a network of wide-field cameras
at three sites around the globe to monitor stars for the presence of
transit signals. The first site in New Mexico has been constructed and
has been taking data since November 2009. The cameras are wide-field in
order to concurrently monitor as many stars as possible. Since a typical
hot Jupiter planet has only a $\sim$10\% probability of transiting the
host star from geometrical considerations, and since it has come to
light that $\sim$0.7\% of stars host a hot Jupiter, we expect that $\sim$1 in
1400 dwarf stars will show a $\sim$4 day periodic transit signal. In our
5.3$^\circ$ FoV, we will be observing anywhere from 10000 to
50000 stars simultaneously, the exact number depending on how close we
point towards the Galactic plane.

Three QES sites have been chosen strategically to provide better
temporal coverage of northern and equatorial stars when combined with
the SuperWASP data in future cooperation on chosen fields. The ``New
Mexico Skies'' observing station, located in southern New Mexico at
latitiude~+32$^\circ$54'14", longitude~105$^\circ$31'44" and
elevation 7200 feet, was chosen to complement the SuperWASP-North
telescope on La Palma.
Our aim is to deploy similar or improved facilities at two additional
northern sites at complementary longitudes in order to be able to more
rapidly establish ephemerides for transiting exoplanet candidates.

The QES project has developed a customised data pipeline using the
{\tt DanDIA} image subtraction software\footnote{{\tt DanDIA} is
built from the DanIDL library of {\tt IDL} routines available
at http://www.danidl.co.uk} to process the data and an archive
compatible with that of SuperWASP to handle the imaging and light curve
datasets necessary for a project with a data rate that is similar to
SuperWASP, both outlined in the later sections. We expect $\sim$50~Gb of
data per clear night from each site. The data are currently partially
processed on site before being fully reduced by the pipeline software
system and archived. The pipeline processing has so far been performed
at the University of St~Andrews while the archive has been hosted at
the Universities of Leicester and Warwick. Currently we are moving all
processing and archive operations to the Qatar
Environment and Energy Research Institute.

\section{The First-Generation QES Wide-Field Camera System}
\label{hardware}

\begin{figure}[htb]
\includegraphics[width=\textwidth,height=\textheight,keepaspectratio]{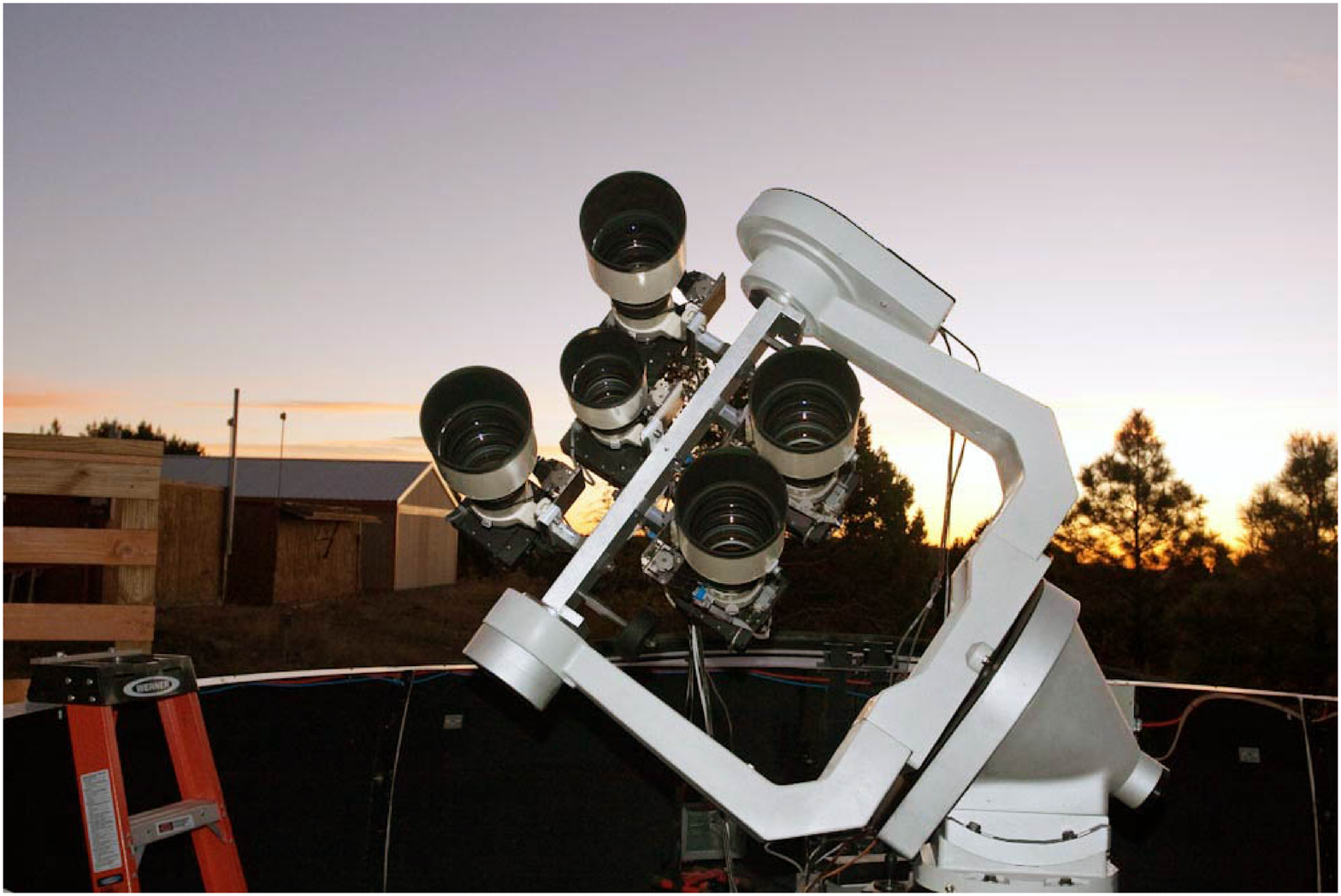}
\FigCap{The first QES observing station, in New Mexico, is fully functioning.
        It consists of four 400mm lenses and one smaller 200mm lens.}
\label{fig2}
\end{figure}

The first QES site, in New Mexico, hosts one wide-field camera system
described here and summarised in Table~\ref{table1}.
The camera system consists of four 400mm
f/2.8 Canon lenses and one 200mm f/2.0 Canon lens each with a FLI
ProLine PL16801 camera with a 4K$\times$4K-pixel KAF-16801E CCD chip.
Each CCD is chilled to a temperature of -40$^\circ$~C to minimize the
dark current inherent in such devices. All five cameras are mounted on a
Mathis equatorial mount MI-750 (Figure~\ref{fig2}). The 400mm lenses
each have a $5.3^\circ\times5.3^\circ$ FoV, a pixel scale of 4.64
arcsec/pixel and cover the magnitude range from 11-15$^{th}$ magnitude
by employing an exposure time of 100~s. The 200mm f/2 lens has an
$11^\circ\times11^\circ$ FoV, a pixel scale of 9.26 arcsec/pixel and
covers the magnitude range from 8-12$^{th}$ mag by employing an exposure
time of 60~s. The FoV of the 200mm lens encapsulates the combined fields
of the four 400mm lenses providing photometry of all stars in the field
in the range from 8-15$^{th}$ magnitude.

\MakeTable{||l|r||}{12.5cm}{First-Generation QES Wide-Field Camera System.}
{\hline
CCD  &  5 KAF-16801E-2 , 4k x 4k \\ \hline
Lenses  & 4x400mm, f/2.8, 1x 200mm, f/2 \\ \hline
Mount  & Equatorial Fork type mount \\ \hline
FoV  &  $5.24^\circ\times5.24^\circ$ per CCD for 400mm lens \\ \hline
Pixel scale &  4.64 arcsec/pix for 400mm lens \\ \hline
Gain &   1.7 e$^-$/ADU \\ \hline
Peak Q.E. &   65\%  \\ \hline
Zero point  & 1 ADU/s at 23 mag \\ \hline
}
\label{table1}

The focus of the camera lenses is very important for data reduction by
difference image analysis (DIA). This is because all DIA algorithms
have problems constructing a kernel solution from images with
under-sampled PSFs. To investigate the focus effects, different hardware
focus setups were used during the first four months of operations.
Data taken from 2009~November to 2010~January with the lenses focused
provided images with an average FWHM of 1.7 pixels. During 2010 February and March
data were taken with the focus set to blur the images to a FWHM of $\sim$3.5 pixels.
Comparing the data from the two campaigns showed that the out-of-focus data gave
better results than the in-focus data when the out-of-focus data had a FWHM of close
to the target value. However, it was found that by defocusing the lenses the FWHM became
very variable and sometimes produced donut-shaped PSFs. As this
resulted in poor photometry, we decided to focus the lenses until
hardware upgrades could be made to automatically stabilise the out of
focus data.

The data acquisition system (DAS), for automatic scheduling and image
acquisition, consists of locally networked Windows PC's with one PC
assigned to each CCD camera. A master PC provides control of
the mount and synchronises the remaining slave PCs, coordinating slews,
focusing, calibration frames as well as when science frames are taken.
For example, once a slew is completed as instructed by the master, the
master begins its exposure and instructs the slaves to begin theirs.

CCDAutoPilot version 4 is a proprietary commercial product that is used
extensively by the amateur astronomy community. It is not open source.
It was modified by the author (Smith) to provide continuous operation,
master/slave control of the multiple systems, coordinated data and
calibration frames including sky flats, simultaneous focusing and
appropriate file and folder naming for compatibility with SuperWASP.
It acts as an executive program controlling other
programs for mount slewing, camera operation and focusing via the
Windows ActiveX interface. Other software used is TheSky6 by Software
Bisque for telescope control, MaxIm DL by Diffraction Ltd for camera
control, FocusMax by Larry Weber and Steve Brady for focusing, and
PinPoint by DC-3 Dreams for plate solving and WCS insertion. An instance
of CCDAutoPilot runs on each PC and has been modified to run
continuously, night after night, without operator intervention as well
as for compatibility with the pipeline-processing program. The software
handles weather interrupts by idling the system until the weather
clears. If the dome is closed due to adverse weather, the system
continues to idle so that it can continue data acquisition once the
weather clears and the dome is opened.

For certain session phases, the slaves are autonomous. For example, when
sky flats are initiated, all systems determine their own exposures to
achieve the targeted signal level via an automatic exposure routine. In
a similar manner the desired number of dark and bias frames
are acquired.  When it is desired to update the focus, each system is
instructed to run an automatic focus routine. After an activity is
instructed to begin, all systems report back when they have completed
that activity. When all have reported an idle condition, the systems are
instructed to begin the next activity.

The target list is specified by a simple text file and is defined for
the year. Another simple text file defines the base exposure time and
cooling temperature for each camera. These text files can be accessed by
non-Windows PC's, thus not requiring direct access to the DAS. Each
evening the software determines the target to be used, based on its
elevation and proximity to the moon. When that target sinks to a
specified elevation in the west, another target is chosen by the same
selection process. WCS coordinates are inserted into the FITS header of
the science frames at the end of the evening's session and the data are
subsequently presented for transfer by the pipeline.

\section{QES Observing Strategy}

\begin{figure}[htb]
\includegraphics[width=0.5\textwidth,height=0.5\textheight,keepaspectratio]{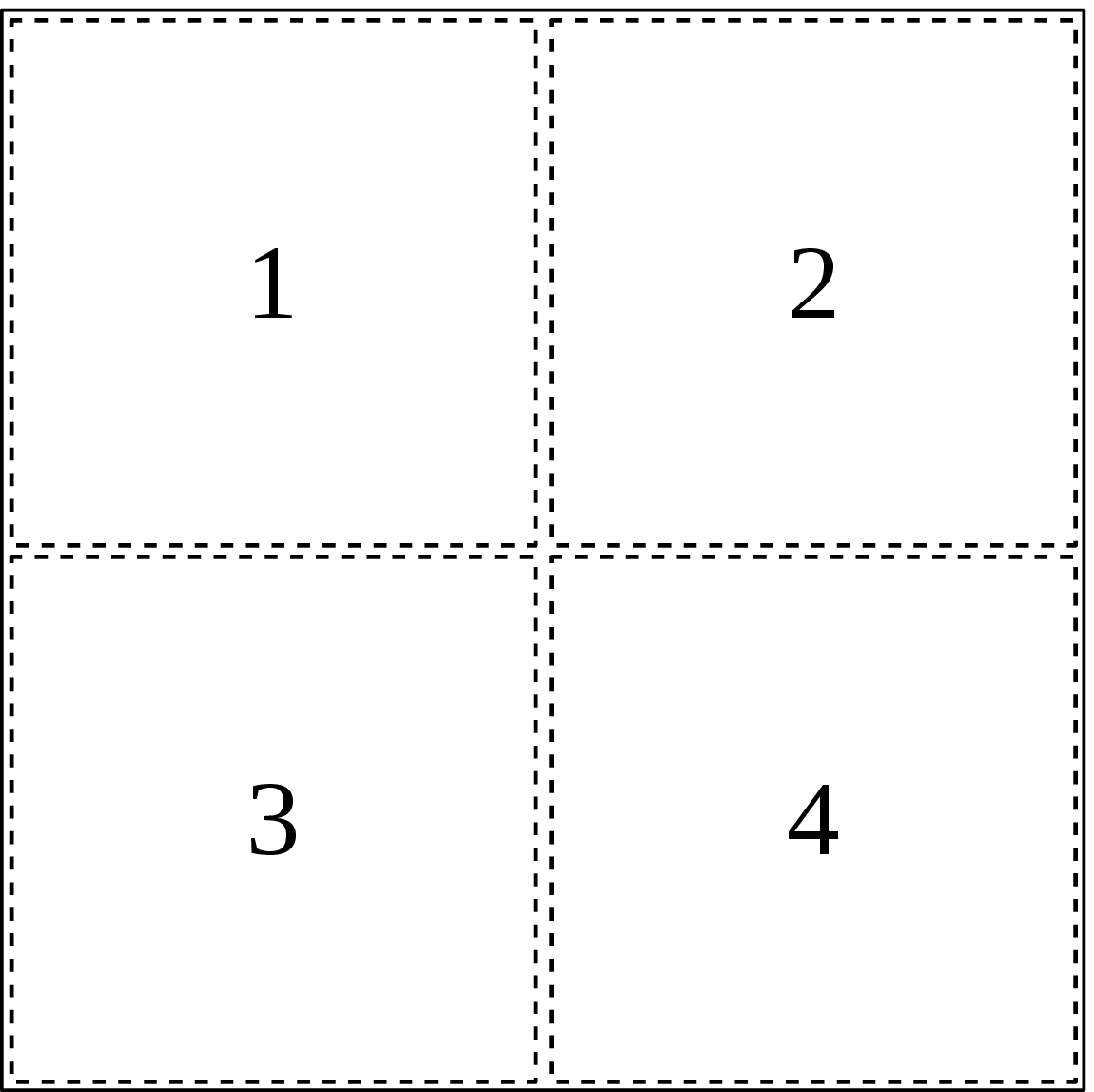}
\includegraphics[width=0.5\textwidth,height=0.5\textheight,keepaspectratio]{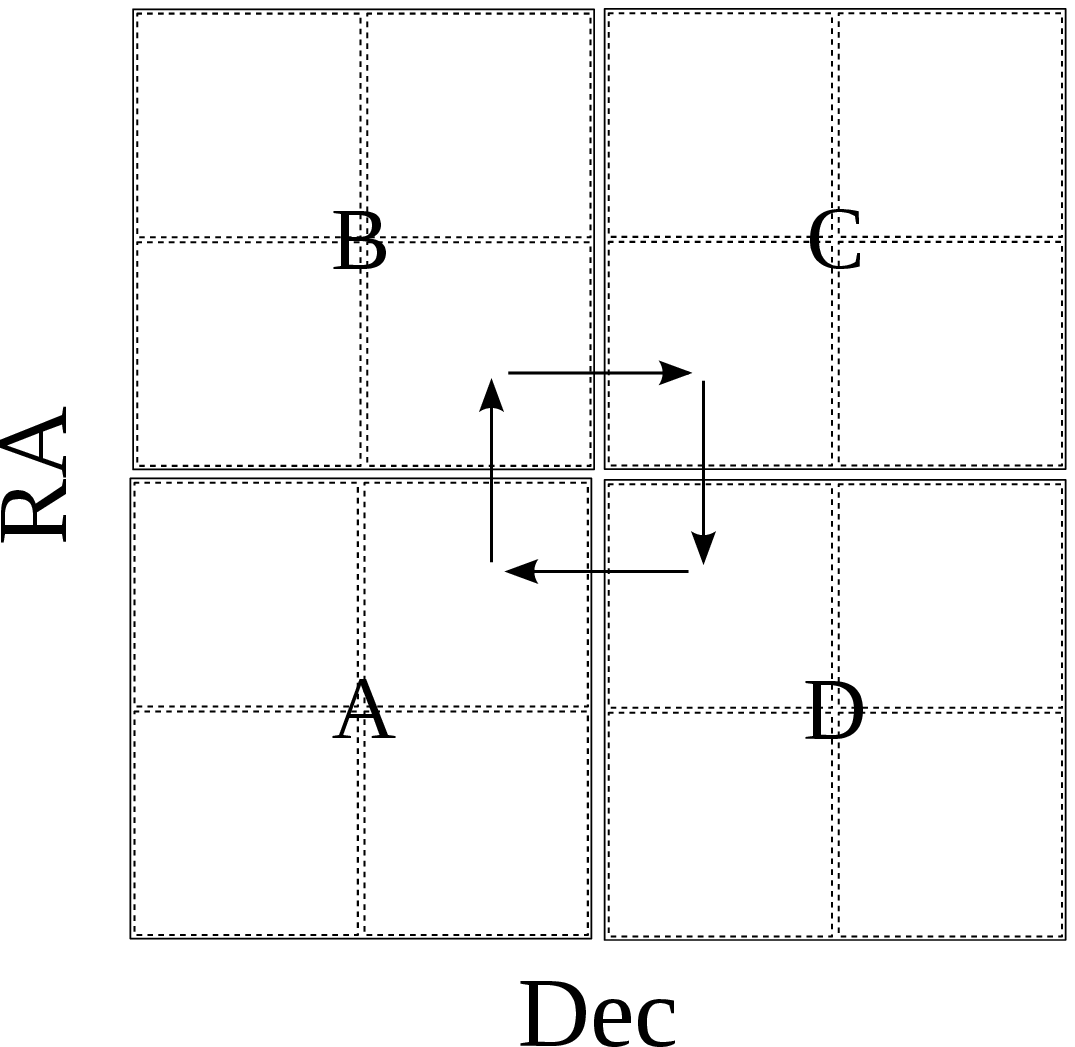}
\FigCap{The 5.3$^{\circ}\times$5.3$^{\circ}$ fields of view of the 400mm lenses
        are delineated by the smaller boxes (dotted squares) labelled 1 through 4.
        The 11$^{\circ}\times$11$^{\circ}$ field of view of the 200mm lens is
        delineated by the larger enclosing box (solid square). The whole system
        moves from pointing A to pointing D continuously all night to cover
        in total $\sim$400 square degrees.}
\label{fig3}
\end{figure}

A full field is defined as a $2\times2$ mosaic of four sub-fields,
numbered 1, 2, 3 and 4 (Figure~\ref{fig3}). Each sub-field is covered
by the FoV of one of the four 400mm lenses,
and all four are covered by the encompassing FoV of the 200mm lens.
While the 400mm cameras take synchronised 100~s exposures,
the 200mm camera takes a 60~s exposure. With these
exposure times the 200mm lens records brighter stars
from 9 to 12.5 visual magnitude with an RMS accuracy of 1\%,
while the 400mm lenses reach deeper to fainter stars
from 10 to 13.75 visual magnitude.

During readout of the CCD images from the 400mm cameras, the mount slews
from pointing A to B (Figure~\ref{fig3}). Once there, the 400mm cameras
again take 100~s exposures while the 200mm camera takes
a 60~s exposure. The slew and exposure sequence then repeats,
moving from pointing B to C, and then to pointing D.
This full cycle of 4 pointings takes approximately 8 minutes to
complete, including the 20~s CCD readout times.
Faint stars in the 12-15$^{th}$ mag range have a photometric measurement
from the 400mm camera system, brighter 8-10$^{th}$ mag stars have a
photometric measurement from the 200mm camera,
and stars in the intermediate range 11-12$^{th}$ mag are
recorded by both the 200mm and 400mm cameras.
Thus every 8 minutes a sky area of $\sim$400 square degrees
is recorded by both the 200mm and 400mm cameras.

The cycle of measurements is repeated at an observing site throughout
the night so long as the chosen field remains higher than 30 degrees
above the horizon, below which the airmass
is too high for reliable wide-field photometry.
A field setting at one site can be picked up by the next site to the
west, providing a capability for continuous coverage
of the field apart from occasional breaks due to bad weather.
As we are aiming to capture the transits of planets with periods from $\sim$0.5-10
days, we observe each field for a period of 2 months before moving
to the next field.  In one year, we can cover up to 10 fields.

\section{QES Data Reduction}

We employ difference image analysis based on the {\tt DanDIA} software to
achieve optimal extraction of light curves from the QES images. The
pipeline has a control program to make it fully automatic. The
difference image analysis procedure begins by selecting the sharpest
(best seeing) image as the reference image.  Constraints are imposed on
sky background and sky transparency to prevent selecting a reference
image with a sharp focus but poor signal-to-noise ratio due to thin
cloud and/or bright moon-lit sky background. For each detected star, the
flux measured on the reference image (referred to as the reference flux,
$f_0$) is found by optimal scaling of the star's point-spread function
(PSF) to fit the reference image data in a pixel box centered on the
star's position. The PSF varies considerably over the wide field of the
QES images, and we account for this by using a spatially-variable
empirical point spread function (PSF) model fitted to all of the
detected PSF-like images. Deblending of very close objects is
attempted. Stars are matched between each image in the sequence and the
reference image, and a linear coordinate transformation is derived and
used to resample the images to register them with the star positions on
the reference image.

\begin{figure}[htb]
\includegraphics[width=\textwidth,height=\textheight,keepaspectratio]{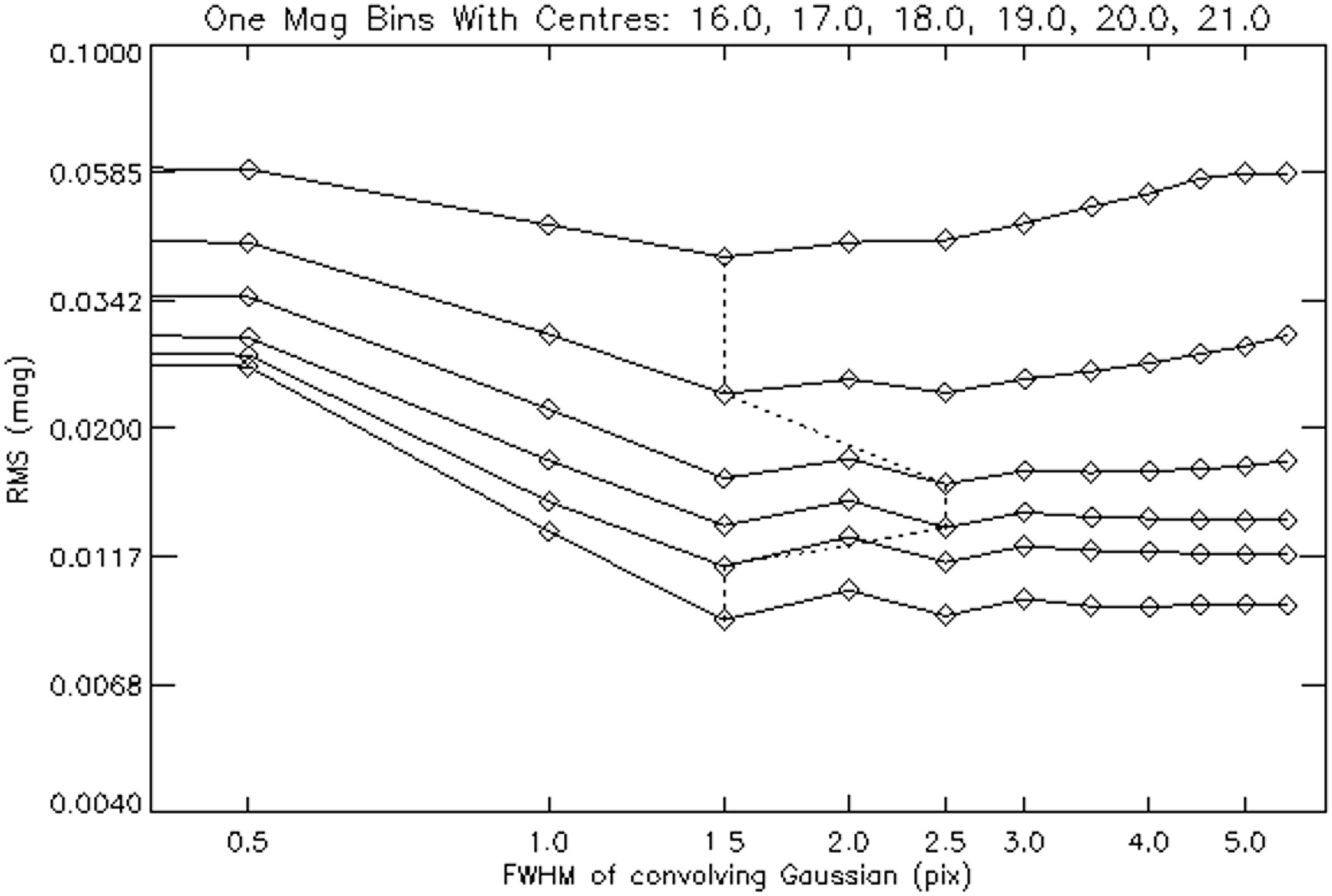}
\FigCap{The effect of different convolving Gaussians used to blur in-focus images,
        with FWHM of $\sim$1.7~pixels, on the light curve RMS scatter once the
        data has been processed by difference image analysis. The figure shows
        the effect of the different Gaussian blurs for 6 different instrumental
        magnitude bins corresponding roughly to V mag 11, 12, 13, 14, 15 and 16
        respectively.}
\label{fig4}
\end{figure}

As mentioned in Section~\ref{hardware}, in-focus QES images have an
under-sampled PSF and yield poor difference image results. We find that
{\tt DanDIA} gives better photometry on these under-sampled images if we first
convolve the reference and registered images with a Gaussian function
that blurs the star images into adjacent pixels. Figure~\ref{fig4} shows
how the Gaussian function's full-width at half-maximum (FWHM) affects
the light curve RMS for stars in different magnitude bins. From the
figure we see that the best results arise after blurring the images with
a FWHM between 1.5 and 2.5 pixels. Based on these results, we keep the
QES cameras in focus, but blur the reference and registered images
by a Gaussian FWHM of 2.5 pixels before performing image subtraction.

Image subtraction is preformed using the methods from
Bramich (2008). We sub-divide the images into a
$14\times14$ grid of cells and determine for each cell
a kernel function, modelled as a pixel array,
derived by matching the PSF in each cell of the reference
image with the PSF in the corresponding cell of the registered
image. The kernel function for each image pixel is obtained by
bi-linear interpolation in the grid of kernels.
The reference image, convolved with the appropriate kernel function
is then subtracted from each registered image to produce a sequence of
difference images.

The differential flux, $\Delta f$,
for each star detected in the reference image
is measured on each difference image as follows. The empirical PSF at
the measured position of the star on the reference image is determined
by shifting the empirical PSF model corresponding to the nearest pixel
by the appropriate sub-pixel shift using image resampling. The empirical
PSF model is then convolved with the kernel model corresponding to the
star position on the difference image. Finally, the PSF is optimally
scaled to fit the difference image at the star position using pixel
variances following a standard CCD noise model.

As the reference image is convolved and scaled to match the
registered image, the flux change $\Delta f$ needs to be
scaled by the photometric scale factor $p$ derived from the image
subtraction kernels as described in Bramich (2008). To
correct for partial cloud cover and transparency variations across the
large field of view of the lenses, the photometric scale factor is
allowed to be spatially variable. This helps improve the light curve
quality achieved by better modelling the poorer quality data
(Bramich et al. 2013). The photometric scale factor is therefore solved for in
each grid cell, and interpolated in the same way as the kernel and
the differential background.

Light curves for each star are constructed by calculating the total flux
$f(t)=f_0+\Delta f(t)/p(t)$
at time $t$ as the sum of the reference flux $f_0$
and the time-dependent difference flux $\Delta f(t)$,
corrected by the time-dependent (and spatially variable) photometric
scale factor $p(t)$.
Fluxes are then converted to instrumental magnitudes via the standard formula
$m=25-2.5\log(f)$, where $m$ is magnitude and $f$ is flux (ADU/s).

To reduce the number of faint stars with a signal-to-noise
ratio too low for exoplanet detection, the stars detected by the pipeline are
matched with the UCAC3 catalogue (Zacharias et al. 2010) and any
objects that lack a matching UCAC3 star are not passed to the
archive. A magnitude zero point correction is also performed between the
reference magnitudes and the UCAC3 magnitudes for all the stars in a
given field using a global SVD fit. The resulting absolute photometric
correction has a mean RMS scatter of approximately 0.1 magnitudes.

For the first week of observations of any field, the images are simply
calibrated for the standard bias, dark and flat field corrections.
At the end of the first seven nights of clear
weather, the best seeing image with an acceptable sky background is
automatically chosen by the software as the reference image. The best
seeing image is required since this image will be convolution matched to
every other image. The pipeline then produces differential
photometry via difference image analysis as described above for all
images to date, and on a nightly basis for all subsequent observations. The
results of the reductions are automatically uploaded for ingestion into the
database archive.

\section{QES Analysis Strategy and Archive}

The reduced photometric data are stored in a data archive system based
on that developed for SuperWASP and described in detail by
Pollacco et al. (2006). The data from each of the survey fields are
treated to remove trends due to instrumental systematic errors using the
SYSREM algorithm (Tamuz et al. 2005, Collier Cameron et al. 2006) and the
Trend Filtering Algorithm (Kov\'acs et al. 2005).

To search for transit signatures we use an evolution of the
box-least-squared (BLS) detection scheme described by
Collier Cameron et al. (2006). The BLS search covers a period range from
0.35-10.0 days. At present we exclude periods in the ranges 0.53-0.57 and
0.95-1.05 days, as detections in these period ranges are almost
invariably spurious and due to residual instrumental systematic effects.
The BLS code is tuned to search for box-like signatures with durations
in the range 1.5-3.75h. We investigate the performance of the
latest methods (BLS, AoVtr, etc., see Tingley 2003 for a
review) on our data sets, and apply more than one method to
identify the most convincing transit candidates.

Once a candidate transit signature is detected by the BLS code, its
parameters are further refined using a Markov Chain Monte-Carlo (MCMC)
algorithm as described by Collier Cameron et al. (2007). The results from
the MCMC analyses of the candidate transits are then imported into a
database, and subjected to manual filtering, eye-balling and
prioritization before being fed into the follow-up programme.

\section{QES Follow-Up Strategy}

The first stage in the follow-up of convincing transit candidates is to
estimate the stellar density and the planetary radius by fitting the
transit profiles from the survey data themselves. We use the
pre-filtering methodology developed by the WASP Project
(Collier Cameron et al. 2007) to identify candidate planetary systems. The
stellar effective temperature is estimated from the 2MASS $J-H$ colour
index. This yields an estimate of the stellar mass under the assumption
that the star is on the main sequence, and a set of non-linear
limb-darkening coefficients as tabulated by Claret (2000).
We use the small-planet model of Mandel \& Agol (2002) to fit the
transit light curve as function of the epoch $T_{0}$ of mid-transit, the
orbital period $P$, the total duration $t_{T}$ from first to fourth
contact, the ratio $R_p/R_\star$ of the planetary to the stellar radius,
and the impact parameter $b$ of the planet's trajectory across the
stellar disc. A Markov-chain Monte Carlo (MCMC) algorithm is used to
determine the posterior probability distributions for each of the
fitting parameters.

The posterior probability distribution for the planetary radius yields
the probability that the planet has a radius less than 1.5 times that of
Jupiter. We also determine the departure of the fitted stellar radius
from the main-sequence value expected for a star of the catalogued $J-H$
colour. As Sozzetti et al. (2007) have noted, the stellar density is
related in a fundamental and model-independent way to the ratio of the
transit duration $t_{T}$ to the orbital period $P$. The location of the
system in a plot of $R_\star/M_\star^{1/3}$ versus $T_{\rm eff}$ gives a
direct assessment of the star's proximity to the main sequence. Many
astrophysical false-positive configurations in which an eclipsing
stellar binary is blended with a brighter star can be detected and
eliminated because the stellar density derived from the transit duration
is inconsistent with the effective temperature derived from the $J-H$
colour. At this stage we also fit a cosine curve with half the orbital
period to determine the amplitude and significance of any ellipsoidal
variation out of transit. Any significant tidal distortion of the
primary indicates that the orbiting companion must be of stellar mass,
eliminating the system as a planet candidate (Sirko \& Paczy\'nski 2003).

For candidates whose transit parameters indicate an object of
planet-like radius orbiting a star that appears to be on the main
sequence, the next step is to obtain high-accuracy high-cadence
light curves covering the suspected transit event, and covering the light
curve phase where a secondary eclipse might occur if the system is an
eclipsing binary. For the brightness range of our survey target stars,
this can be achieved by a 1m-class telescope. Transit candidates are
rejected upon the detection of secondary eclipses, ellipsoidal
variations, and/or heating effects, all indicative of an eclipsing
binary rather than a bona fide transiting planet. The higher angular
resolution of the 1m-class telescope helps us to resolve cases where the
transits originate in a nearby, faint stellar binary located close to a
brighter star. Further MCMC analysis of the follow-up transit light
curve and host star properties (colour, spectral type etc.) yields a
minimum radius for the transiting body, which can be used to reject a
planetary transit candidate if its value is too large ($> 2 R_{J}$).

For the purpose of follow-up observations, the SuperWASP Alsubai
Follow-up Telescope (SAFT) is being constructed as a 1-m robotic
telescope facility on La Palma in the Canary Islands. The aim of the
telescope is solely for transit candidate follow-up, shared with the
SuperWASP Project. The QES Project will access $\sim$35\% of the
observing time in return for the investment in the construction and
running costs. The project has been granted permission to place the
telescope at the observatory site, and the construction is in progress.
It is estimated to be finished most likely in early 2014.

Candidates that pass the photometric follow-up stage are then placed on
candidate lists for radial-velocity measurements. The existing
collaborations between WASP Consortium and the Geneva CORALIE team
and the French SOPHIE consortium have developed a highly efficient and
successful strategy modelled on that described by
Pont et al. (2005) for Doppler follow-up of OGLE transit
candidates. An initial reconnaissance observation is used to screen for
extreme rotation (which precludes determination of a planetary orbit) or
obvious double-lined spectroscopic binaries. Surviving candidates are
re-observed a day or so later, to eliminate single-lined binaries with
unseen companions of stellar mass. Subsequent radial velocity
observations target the quadrature phases of the orbit, to determine the
total radial-velocity amplitude and hence the planetary mass. Further
observations are then obtained to determine the radial acceleration near
both conjunctions, to estimate the orbital eccentricity. For candidates
brighter than $V=13$~mag or so, we combine our candidate lists with the
WASP programmes on CORALIE in the southern hemisphere and SOPHIE in the
north. For fainter candidates we will submit our own observing proposals
to carry out radial-velocity follow-up using 4-10m-class telescopes.

\section{Initial Results from QES}

\begin{figure}[htb]
\includegraphics[width=\textwidth,height=\textheight,keepaspectratio]{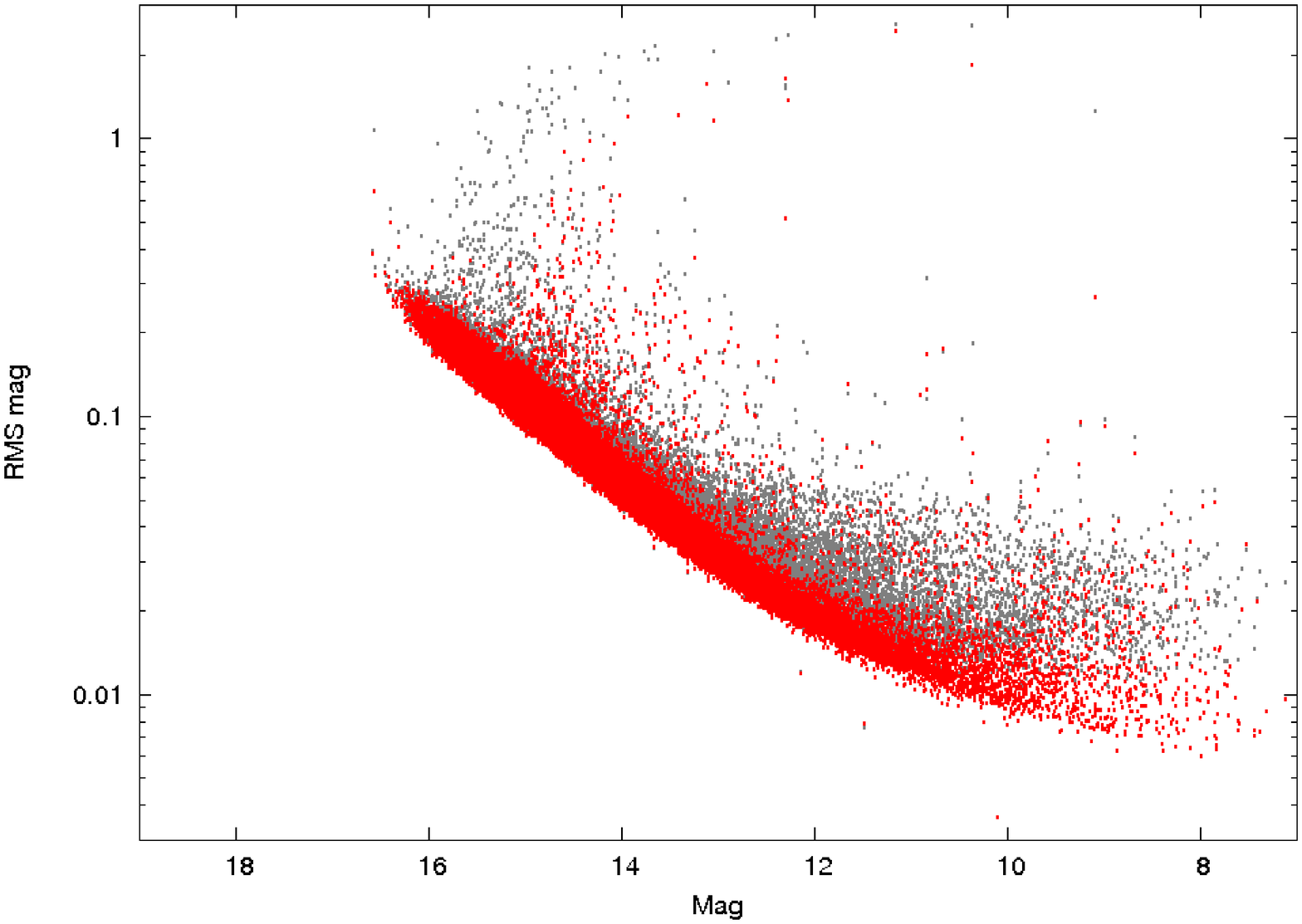}
\FigCap{RMS plot showing the star magnitude residuals for the 200mm camera
        for a whole field's worth of data. The RMS scatter is shown before (grey)
        and after (red) being detrended with the SYSREM algorithm.}
\label{fig5}
\end{figure}

\begin{figure}[htb]
\includegraphics[width=\textwidth,height=\textheight,keepaspectratio]{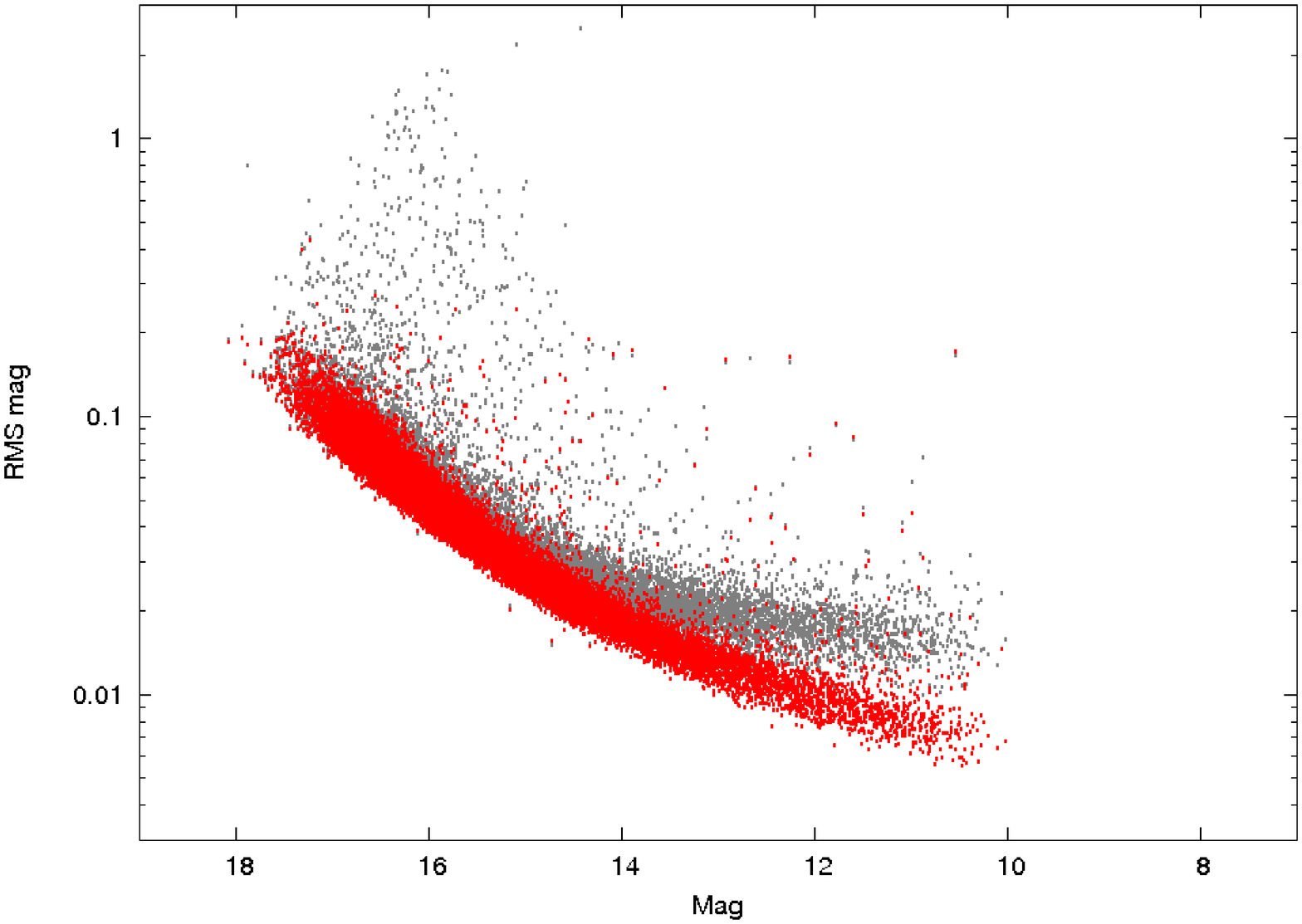}
\FigCap{RMS plot of the star magnitude residuals for one of the 400mm cameras
       	for a whole field's worth of data. The RMS scatter is shown before (grey)
       	and after (red) being detrended with the SYSREM algorithm.}
\label{fig6}
\end{figure}

Figures \ref{fig5} and \ref{fig6} show the RMS of the magnitude
residuals for two fields taken with the 200mm and the 400mm cameras,
respectively. These plots show the residuals before and after being
detrended with the SYSREM algorithm. It can be seen from the two figures
that as planned the 400mm camera samples a much fainter magnitude
range with a good RMS precision. The 200mm camera is imaging the bright
stars and providing an overlap in detection for stars
in the $V$ magnitude range 10 to 12.
Many QES candidates are discovered independently by both the 200mm and
400mm cameras, in the magnitude range where the two systems overlap. This
dual detection capability adds confidence that
the detected transits are real and not due to systematic errors.
The QES data are also producing transiting
candidates around much fainter stars, with some
$V\sim15^{th}$ mag stars being flagged as candidates.

\begin{figure}[htb]
\includegraphics[width=0.5\textwidth,height=0.5\textheight,keepaspectratio]{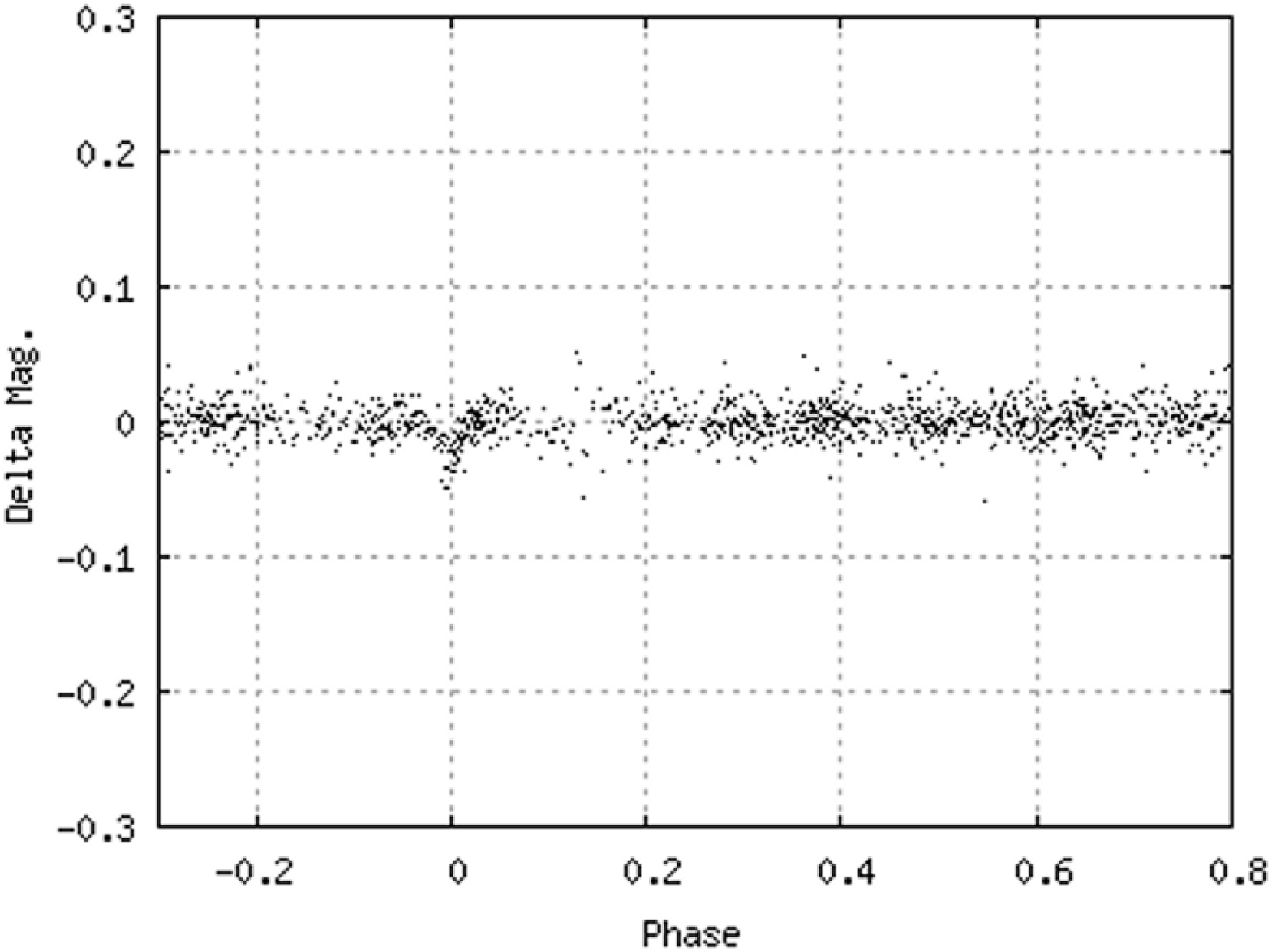}
\includegraphics[width=0.5\textwidth,height=0.5\textheight,keepaspectratio]{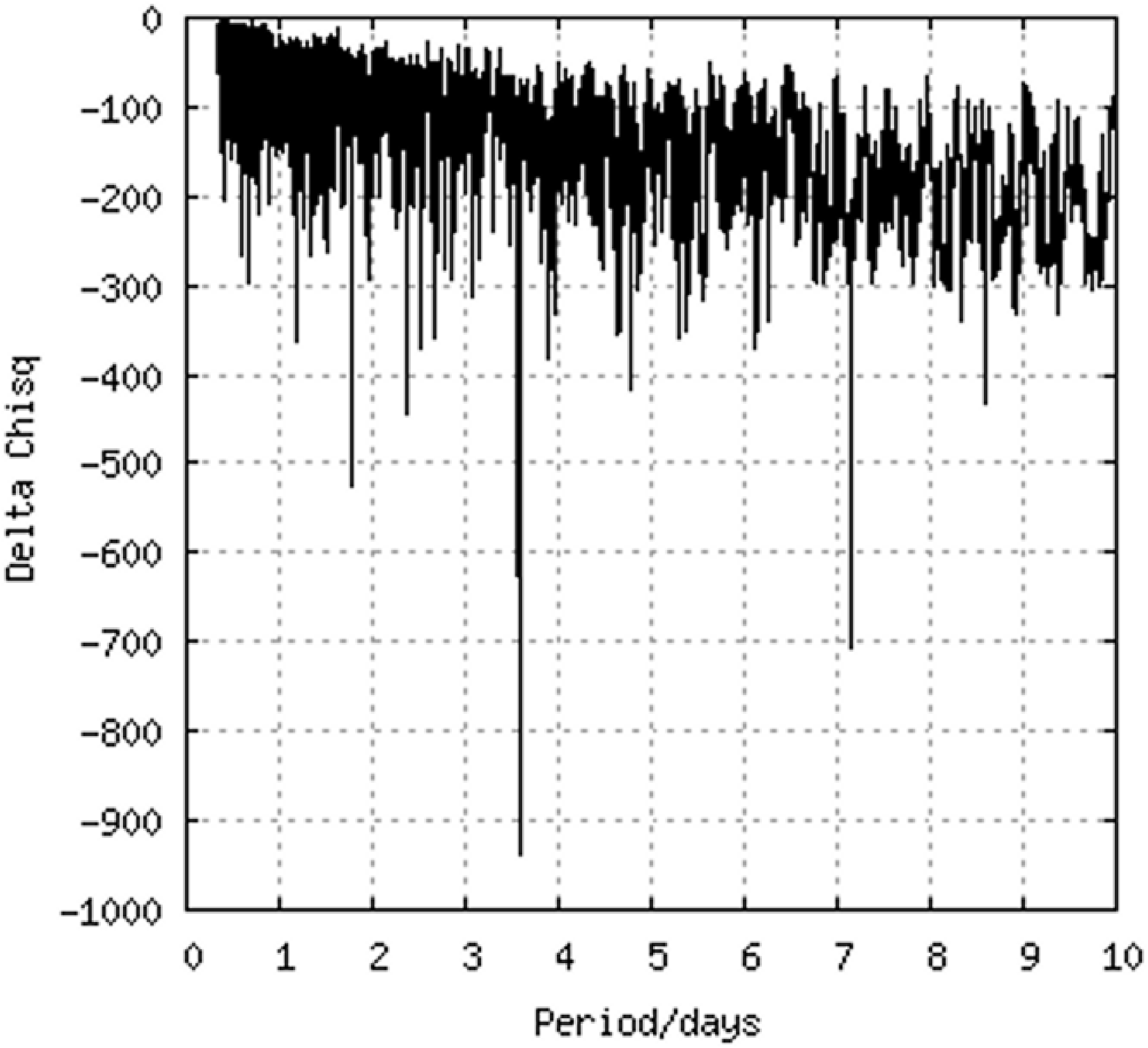}
\FigCap{QES light curve and periodogram of WASP-37b, an exoplanet detected by SuperWASP
        (Simpson et al. 2011) and independently flagged in the QES database.}
\label{fig8}
\end{figure}

\begin{figure}[htb]
\includegraphics[width=0.5\textwidth,height=0.5\textheight,keepaspectratio]{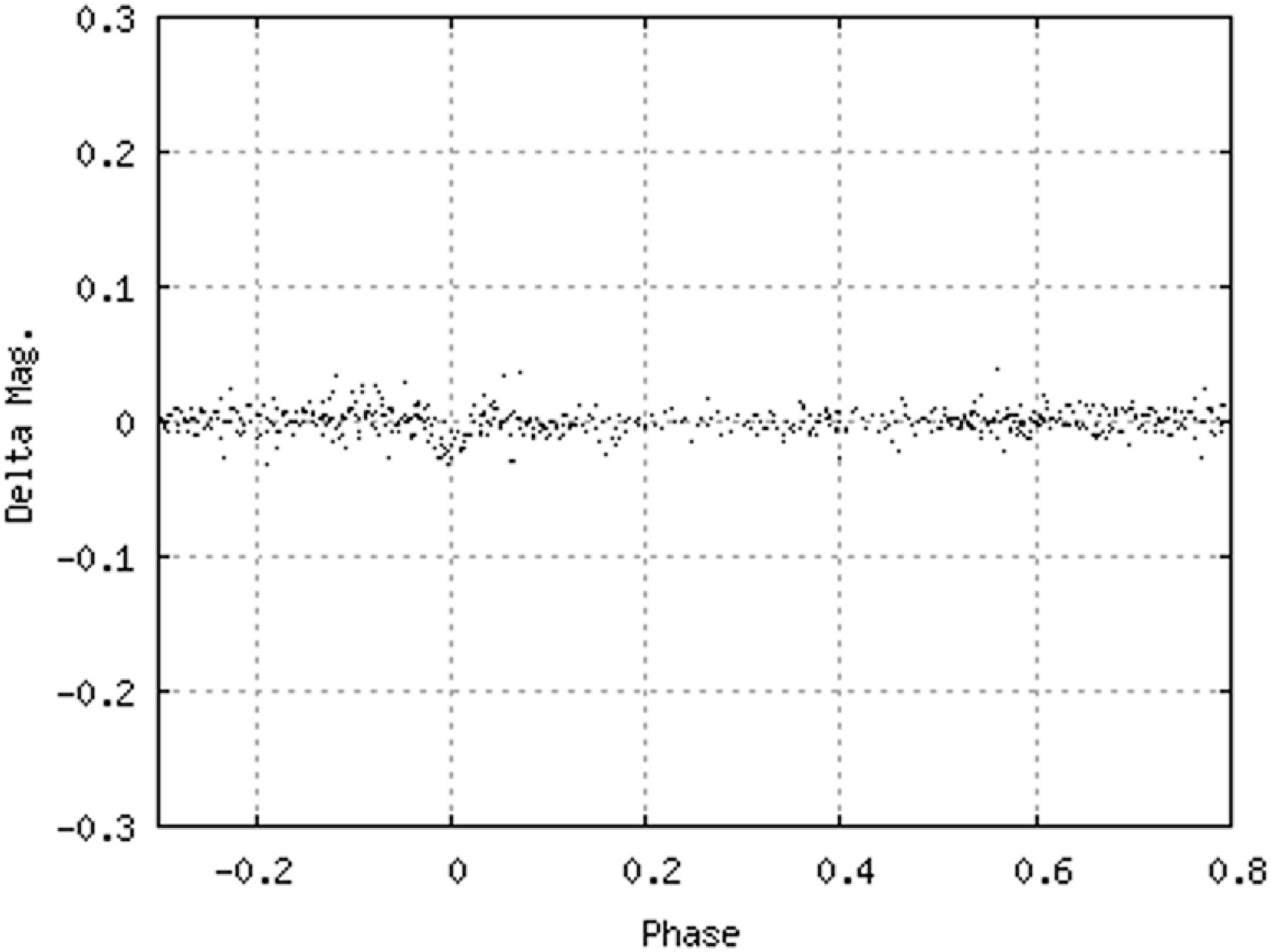}
\includegraphics[width=0.5\textwidth,height=0.5\textheight,keepaspectratio]{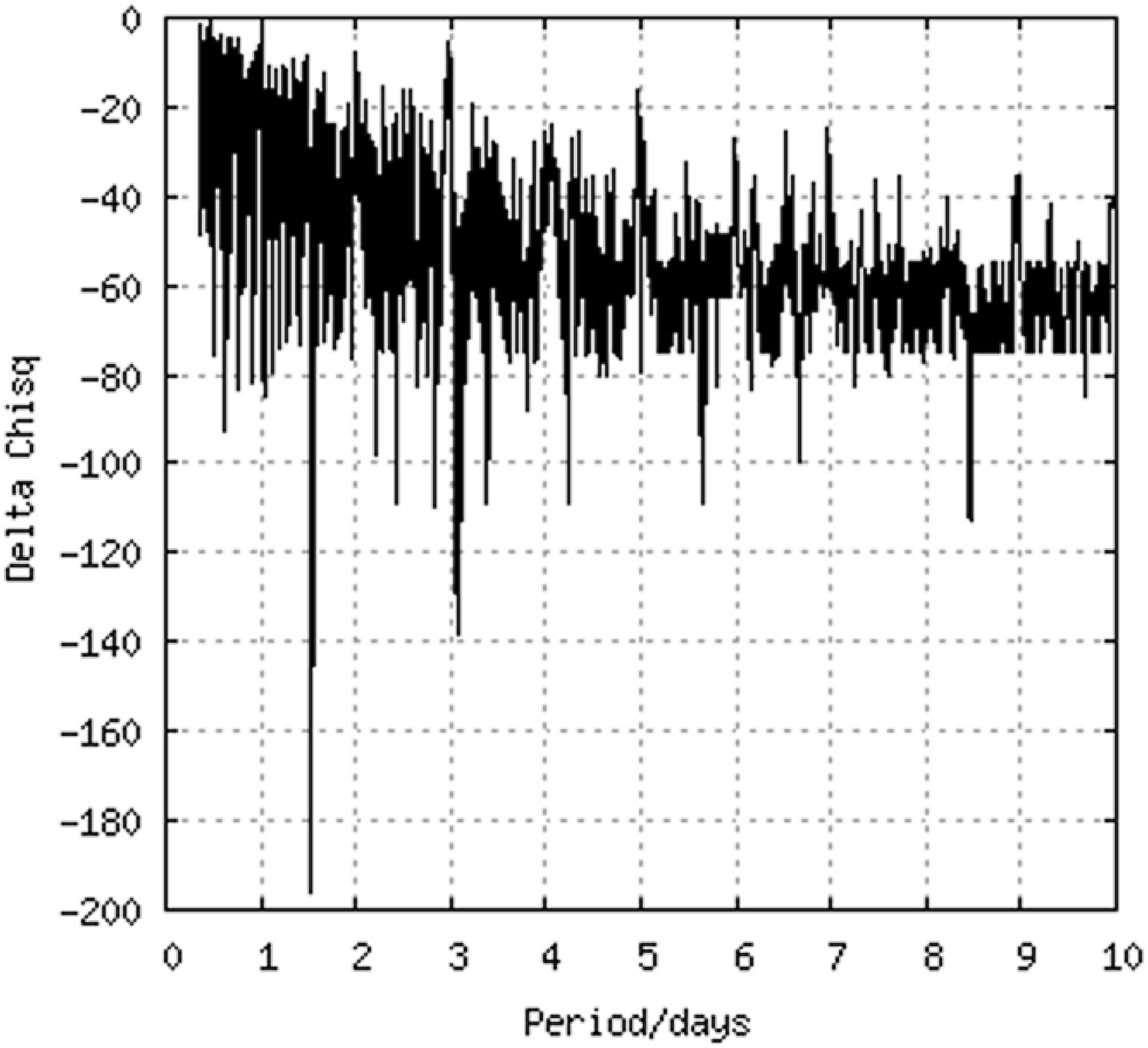}
\FigCap{QES light curve and periodogram of WASP-36b, an exoplanet detected by SuperWASP
        (Smith et al. 2012) and independently flagged in the QES database.}
\label{fig9}
\end{figure}


QES is also independently identifying a number of transiting exoplanets
found first by other surveys such as SuperWASP. Several of these planets
were independently flagged as A-list candidates in the QES data before
later checks with the SuperWASP database showed them to already be
confirmed planets.  Figures~\ref{fig8} and \ref{fig9} show two such
examples. Figure~\ref{fig8} shows the QES light curve and
corresponding periodogram of WASP-37 obtained with one of the 400mm cameras.
The QES data reveal the transit signature and the QES periodogram identifies
the correct period independently of the SuperWASP data that first found the transits.
WASP-37b is a $1.8\,M_J$ $1.2~R_J$ planet
in a 3.58~day orbit around a $V=12.7$~mag metal-poor G2 dwarf
(Simpson et al. 2011).
Figure~\ref{fig9} shows the QES light curve and periodogram for WASP-36.
This $V=12.7$~mag metal-poor G2 dwarf
hosts a $1.3~M_J$ $2.3~R_J$ planet in a 1.54~day orbit
(Smith et al. 2012).
The transit is clearly detected in the QES light curve
from one of the 400mm cameras,
and the QES periodogram identifies the correct
period.

The QES project has been collecting data since mid November 2009.
The first 34 fields that were processed and ingested into the
archive yielded 1,863,375,935 photometric data
points on on a total of 951,417 stars.
QES has identified hundreds of promising candidates,
dozens of which have been promoted to the A-list for
photometric and radial velocity follow up.
The first two QES planets, Qatar~1b (Alsubai et al. 2011)
and Qatar~2b (Bryan et al. 2012), have already been confirmed.

The next stage in the development of QES is a planned deployment
of similar camera systems at two complementary longitudes
in the northern hemisphere.  The more nearly continuous
temporal coverage afforded by a 3-site survey
should greatly reduce the time needed to
identify candidates and establish reliable transit ephemerides
prior to photometric and radial velocity follow-up observations.
Given the quality of the light curves that the QES project is producing
and the effective validation of candidate filtering methods, we
can anticipate that QES will find many more transiting exoplanets
among which will be hot Saturns and hot Neptunes orbiting
stars bright enough for follow-up investigations.

\Acknow{KA, KH, ACC and DMB acknowledge the Qatar Foundation for support from QNRF grant NPRP-09-476-1-078.
        KH is supported by a Royal Society Leverhulme Trust Senior Research Fellowship.}

\end{document}